\documentclass[english,aps,onecolumn,amsmath,amssymb,groupedaddress,showkeys]{revtex4-2}

\usepackage[T1]{fontenc}
\usepackage[utf8]{inputenc}
\usepackage{babel}

\usepackage{amsmath}
\usepackage{mathrsfs}
\usepackage{graphicx}
\usepackage{txfonts}

\DeclareMathOperator{\arcsinh}{arcsinh}

\begin{document}

\title{Ising Model on Calandre Lattice}

\author{F. A. Kassan-Ogly}
\author{A. V. Zarubin}
\email{Alexander.Zarubin@imp.uran.ru}
\address{M. N. Mikheev Institute of Metal Physics of Ural Branch of Russian Academy of Sciences, S.~Kovalevskoy Street 18, 620108 Ekaterinburg, Russia}

\begin{abstract}
In this paper, we proposed to consideration a noval (not yet exploited) planar lattice --- Calandre lattice with the Ising model applied to it. In the framework of an exact analytical Kramers--Wannier transfer-matrix approach and using an extension of the combinatorial Vdovichenko--Feynman method, we have obtained exact solution for Helmholtz free energy and consequent expression for the principal eigenvalue of the Kramers--Wannier transfer-matrix. In addition, we derived the exact expression for spontaneous magnetization. The usage of both expressions made it possible to perform the comprehensive studies of thermodynamic and magnetic properties of the Calandre lattice itself and a wide variety of its affiliated lattices. Among them, diversified interrelations between Calandre lattice and already known and even some novel lattices. These lattices have also revealed such effects as magnetic frustrations, multiple phase transitions and magnetic reentrancy.
\end{abstract}

\keywords{Calandre lattice; Ising model; exact solution; magnetic frustration; magnetic reentrancy} 

\maketitle

\section{Introduction}

To the great detriment of humanity, many remarkable achievements of the Hellenistic era have not survived to the present day. Among them is a lost work by Archimedes, which, in particular, derived planar lattices. Fortunately, this work was accessible to Kepler, who subsequently presented and enumerated the eleven Archimedean lattices in his own book \cite{Kepler:1619}. A proof that these are the only vertex-transitive regular lattices (tilings) is given in Gr\"{u}nbaum's and Shephard's book \cite{Grunbaum:1986}.

For a lattice to be classified as Archimedean, it must satisfy the following necessary and sufficient conditions: 1) The plane must be tiled exclusively by regular polygons without gaps or overlaps. 2) The lattice must be vertex-transitive in the language of graph theory (or translationally invariant with respect to its nodes in the language of crystallography). 3) All vertices (nodes) must be equivalent. A regular polygon is one in which all side lengths are equal and all interior angles are equal as well.

Given these conditions, the Ising model has proven applicable to all Archimedean lattices, allowing for exact solutions for the Helmholtz free energy, similar to the Onsager solution. The outstanding work of Onsager \cite{Onsager:1944}, in which for the first time an exact solution for the Helmholtz free energy on a square lattice in the Ising model was obtained, opened a new era and actually defined the boundary of demarcation between the previous naive and very simplified theories of the mean (molecular) field type and the modern era of the study of critical phenomena in statistical physics with accompanying features behavior of thermodynamic quantities. The most significant consequence of the exact solution is the logarithmic (Onsager's) singularity of the heat capacity near the phase-transition point, observed in all subsequent exact solutions on all planar lattices. 

Due to this circumstance, Archimedean lattices have been drawn into the remarkable world of exact solutions in statistical physics and are currently experiencing their own Renaissance. 

It should be emphasized that Kramers and Wannier \cite{Kramers:1941:2} introduced the concept of the dual lattice (the reciprocal lattice in crystallographic terminology) and, by exploiting the self-duality of the square lattice, derived an exact expression for the phase-transition point three years prior to Onsager's exact solution. Subsequently, this concept began to play a significant role in statistical physics; in particular, dual lattices were derived for all Archimedean lattices---known as Laves (or Catalan) lattices---which, following the Archimedean ones, have since also been involved into the world of exact solutions within the framework of the Ising model.

To date, ever-new lattices that belong neither to the Archimedean nor to the Laves class have emerged and continue to emerge, for which the Ising model is applied and investigated. Among them are the two so-called 4-6 lattices introduced by Oitmaa and Keppert \cite{Oitmaa:2002}; the self-dual 3-5 and 3-6 lattices by Chao et al. \cite{Chao:1990}; various bow-tie lattices \cite{VeraStimpson:2018,Strecka:2008,Zhang:2010,Ding:2012,Scullard:2013} by Vera Stimpson et al.; Stre\v{c}ka and \v{C}anov\'{a}; and Ding et al.; the rectangular-triangular Bose lattice \cite{Bose:1986}; square-kagome lattice \cite{Gembe:2023}; and others.

However, it should be noted that exact solutions have not yet been obtained---and remain to be found---for all lattices, including both Archimedean and Laves ones.

In our paper, a new lattice---also belonging neither to the Archimedean nor to the Laves class---is proposed and studied in detail.

First, in this paper we confined ourselves to investigating Calandre lattice with five exchange interactions mainly with absolute values equal to unity and its dual lattice. Second, we focused our studies on finding all possible particular limiting cases with zero interactions. Third, deriving the exact expressions for phase-transition temperatures, both Curie and N\'{e}el, and for corresponding residual entropies, that gave us opportunity to ascertain various interrelations between Calandre lattice and many already known or noval yet unexplored lattices.

\section{Materials and methods}

\begin{figure}[tbh]
\centering \includegraphics{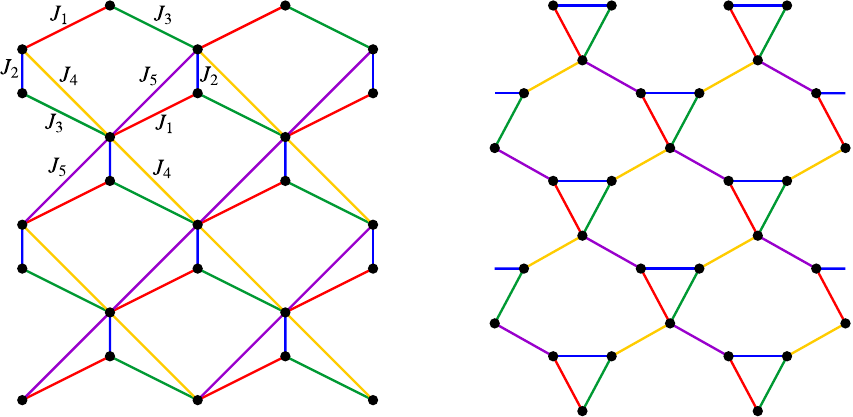}
\caption{The Calandre lattice (left panel) and its dual lattice (right panel)}
\label{fig:01}
\end{figure}

This publication introduces a new lattice, termed Calandre lattice, into scientific discourse. It is a planar, non-Archimedean lattice that exhibits periodicity in two dimensions. A schematic representation of this lattice and its dual lattice is provided in Fig. \ref{fig:01}.

We will conduct further investigation of this lattice within the framework of the Ising model---a fundamental model in statistical physics, traditionally used to describe phase transitions and cooperative phenomena in systems of interacting spins (see, e.g., \cite{Ising:1925,Onsager:1944}). The Hamiltonian of this model, defined on the lattice under consideration, is given by
\begin{equation}
\mathscr{H}=-\sum_{i=1}^{5}J_{i}\sum_{\langle m,n\rangle_{i}}\sigma_{m}\sigma_{n},\label{eq:H}
\end{equation}
here, $J_{i}$ denotes the exchange interaction between atomic spins located at nearest-neighbor lattice sites. The lattice contains five distinct exchange interactions, which are indicated by different colors in Fig. \ref{fig:01}. Each symbol $\sigma$ takes one of two values, $\pm1$, corresponding to a classical description of a spin variable. The notation $\langle m,n\rangle_{i}$ signifies a summation over all pairs of nearest neighbors. Note that the sign of the exchange interaction parameter determines the nature of the magnetic ordering: the condition $J_{i}>0$ corresponds to ferromagnetic ordering, whereas $J_{i}<0$ corresponds to antiferromagnetic ordering.

For the Ising model under consideration on the Calandre lattice, we have obtained an exact solution for the Helmholtz free energy per spin
\[
F=-T\ln\lambda,
\]
employing the Kramers--Wannier transfer-matrix method \cite{Kramers:1941:1,Kramers:1941:2} and an extension of the Vdovichenko--Feynman combinatorial method \cite{Vdovichenko:1965,Feynman:1972}.

It is known that the free energy is a function of the principal eigenvalue (the unique maximal real eigenvalue, as guaranteed by the Perron--Frobenius theorem \cite{Baxter:1982}) of the Kramers--Wannier transfer matrix. In this case, the expression for the principal eigenvalue of the transfer matrix ($\lambda$) can be written for
\begin{equation}
\ln\frac{\lambda}{2}=\frac{1}{16\pi^{2}}\intop_{0}^{2\pi}\intop_{0}^{2\pi}L(\omega,\omega^{\prime})\,d\omega\,d\omega^{\prime},\label{eq:L1}
\end{equation}
\begin{multline*}
L(\phi_{x},\phi_{y})=\ln[\tfrac{1}{2}(C_{1}C_{2}C_{3}C_{4}C_{5}+C_{1}S_{2}S_{3}C_{4}S_{5}+S_{1}S_{2}C_{3}S_{4}C_{5}+S_{1}C_{2}S_{3}S_{4}S_{5}+C_{4}C_{5}\\
-(C_{3}S_{4}+C_{1}C_{2}S_{4}+S_{1}S_{2}C_{4})\cos\omega-(C_{1}S_{5}+C_{2}C_{3}S_{5}+S_{2}S_{3}C_{5})\cos\omega^{\prime}-S_{1}S_{3}\cos(\omega-\omega^{\prime}))],
\end{multline*}
where
\[
C_{i}=\cosh\frac{2J_{i}}{T},\quad S_{i}=\sinh\frac{2J_{i}}{T},\quad i=1,\ 2,\ 3,\ 4,\ 5,
\]
$T$ is the absolute temperature.

It should be noted that in the calculations presented here, the Boltzmann constant is set to unity ($k_{\text{B}}=1$), which corresponds to a system of units commonly used in statistical physics, where temperature is expressed in energy units. Furthermore, to maintain a consistent scale, both temperature and the parameters of exchange interactions are measured in absolute units of exchange interaction. This choice of units ensures the internal consistency of all quantities involved in the calculations and allows for the comparison of results without the need for additional conversions.

Knowing the principal eigenvalue of the Kramers--Wannier transfer matrix (\ref{eq:L1}), one can calculate the thermodynamic functions of the system, including the specific entropy
\begin{equation}
S=\ln\lambda+\frac{T}{\lambda}\frac{\partial\lambda}{\partial T},\label{eq:S}
\end{equation}
and magnetic specific heat capacity
\begin{equation}
C=2\frac{T}{\lambda}\frac{\partial\lambda}{\partial T}+\frac{T^{2}}{\lambda}\frac{\partial^{2}\lambda}{\partial T^{2}}-\frac{T^{2}}{\lambda^{2}}\left(\frac{\partial\lambda}{\partial T}\right)^{2}.\label{eq:C}
\end{equation}

Furthermore, we have obtained an exact analytical expression for the spontaneous magnetization---a quantity defined \cite{Montroll:1963,Baxter:2011} as the square root of the spin-spin correlation function in the limit where the spins are infinitely far apart
\[
M=\sqrt{\lim_{\Delta\to\infty}\langle\sigma_{1}\sigma_{1+\Delta}\rangle}.
\]
The expression sought for the Calandre lattice is as follows:
\begin{equation}
M=(1-k)^{1/8},\label{eq:M}
\end{equation}
where
\[
k=\frac{(C_{1}C_{3}+C_{2})^{2}-(S_{1}S_{3})^{2}}{(C_{1}C_{2}C_{3}S_{4}S_{5}+S_{1}C_{2}S_{3}C_{4}C_{5}+C_{1}S_{2}S_{3}S_{4}C_{5}+S_{1}S_{2}C_{3}C_{4}S_{5}+S_{4}S_{5})^{2}-(S_{1}S_{3})^{2}}.
\]

The thermodynamic and magnetic functions obtained enable a detailed investigation of the magnetic behavior of the Ising system under consideration. This approach, which traces back to the classic works on the statistical theory of the Ising model \cite{Peierls:1936,Onsager:1944,Wannier:1950,Newell:1953,Brush:1967}, remains a productive tool for studying the magnetism of spin systems; it allows for the derivation of quantitative estimates of observable quantities that can be compared with experimental data and numerical calculations.

\section{Results}

\subsection{Curie and N\'{e}el temperatures for the Calandre lattice}

Let us consider an Ising ferromagnet with model parameters (\ref{eq:H}) equal to
\begin{equation}
J_{1}=J_{2}=J_{3}=J_{4}=J_{5}=J=1\label{eq:J:FM}
\end{equation}
and an Ising antiferromagnet
\begin{equation}
J_{1}=J_{2}=J_{3}=J_{4}=J_{5}=J=-1.\label{eq:J:AFM}
\end{equation}
Using formula (\ref{eq:L1}) for the principal eigenvalue of the Kramers--Wannier transfer matrix, and after not very complicated but tedious calculations, we have obtained the following exact expressions
\begin{equation}
T_{\text{C}}=\frac{2}{\ln\left({\displaystyle \frac{1}{3}+\frac{1}{3}\sqrt[3]{19+3\sqrt{33}}+\frac{1}{3}\sqrt[3]{19-3\sqrt{33}}}\right)}\approx3.282\,036\label{eq:TC:Curie}
\end{equation}
for phase-transition temperatures---both the Curie point and 
\begin{equation}
T_{\text{N}}=\frac{2}{\ln\left({\displaystyle \frac{1}{3}+\frac{1}{3}\sqrt[3]{27+3\sqrt{57}}+\frac{1}{3}\sqrt[3]{27-3\sqrt{57}}}\right)}\approx1.963\,495\label{eq:TC:Neel}
\end{equation}
the N\'{e}el point---in the Ising model on the Calandre lattice. Inter alia, it turned out that similar derivation of them by formula (\ref{eq:M}) for spontaneous magnetization was much more easier and gave the very same expressions which give evidence of full consistency between obtained formulas for the Kramers--Wannier transfer-matrix principal eigenvalue and spontaneous magnetization.

\begin{figure}[tbh]
\centering \includegraphics{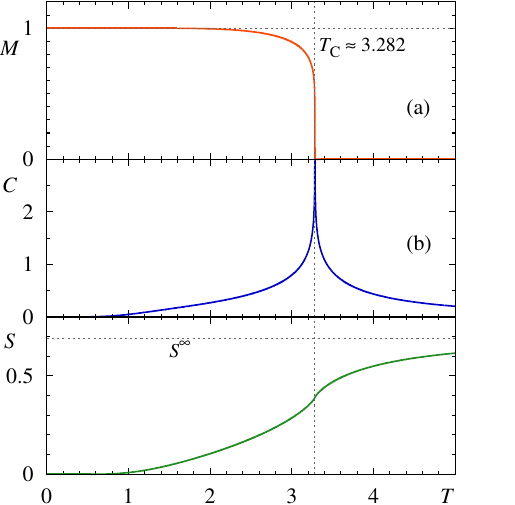}\qquad\includegraphics{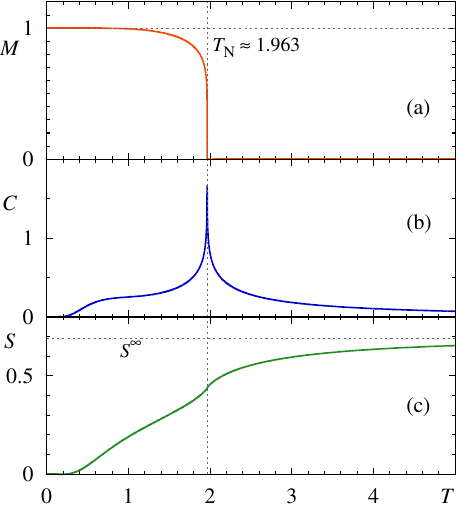}
\caption{Temperature dependencies of spontaneous magnetization $M(T)$ (a), specific heat $C(T)$ (b), and entropy $S(T)$ (c) for an Ising ferromagnet (\ref{eq:J:FM}) (left panel) and an Ising antiferromagnet (\ref{eq:J:AFM}) (right panel) for the Calandre lattice. The vertical dashed line indicates the position of the phase-transition temperature. In panel (c), the horizontal dashed line corresponds to the entropy $S^{\infty}=\ln2$ at $T\to\infty$}
\label{fig:02}
\end{figure}

The temperature dependencies of the thermodynamic and magnetic functions of the spin system exhibit a single magnetic phase transition, corresponding to the points (\ref{eq:TC:Curie}) and (\ref{eq:TC:Neel}); these dependencies are shown in Fig. \ref{fig:02}.

As a next step, we obtained an expression for the phase-transition point on the dual Calandre lattice. This lattice, as yet, does not have an exact solution for the Helmholtz free energy, let alone for spontaneous magnetization. To achieve this, we directly applied our general formula 
\begin{equation}
T_{\text{c}}^{\text{dual}}=\frac{2}{\arcsinh\left({\displaystyle \frac{1}{\sinh{\displaystyle \frac{2}{T_{\text{c}}}}}}\right)},\label{eq:TC:Dual}
\end{equation}
which connects phase-transition points between any given lattice and its dual.

It is worth noting that Kramers--Wannier duality---a distinctive feature of the Ising model on low-dimensional lattices---makes it possible, in principle, to construct a dual lattice for any given planar lattice; consequently, an exact solution known for the original lattice allows one to obtain an exact solution for the dual lattice as well.

At this stage, a truly astounding property of these two ferromagnetic lattices, coupled by the duality relation, was discovered. This property was absolutely unexpected by us, inherent only to them, and not observed in any other known lattices. Specifically, the ratio of phase-transition points is such that on the dual lattice, the phase-transition point
\begin{equation}
T_{\text{C}}^{\text{dual}}=\frac{2}{\ln\left({\displaystyle 1+\frac{1}{3}\sqrt[3]{54+6\sqrt{33}}+\frac{1}{3}\sqrt[3]{54-6\sqrt{33}}}\right)}\approx1.641\,018\label{eq:TC:1.641}
\end{equation}
is exactly half that of the original Calandre lattice,
\[
\frac{T_{\text{C}}}{T_{\text{C}}^{\text{dual}}}=2.
\]

Our further studies of the lattice, dual to Calandre, have revealed that in fully antiferromagnetic case (\ref{eq:J:AFM}) it is frustrated without phase transition and its residual entropy (a non-zero zero-temperature entropy of the system) has the form
\begin{equation}
S^{\circ}=\frac{1}{6\pi}\intop_{0}^{\pi}\ln\left(\frac{1+4\cos^{2}\alpha+\sqrt{\left(1-4\cos^{2}\alpha\right)^{2}}}{2}\right)\,d\alpha\approx0.107\,689.\label{eq:S0:0.108}
\end{equation}
The residual entropy of a fully antiferromagnetic triangular lattice, as derived by Wannier \cite{Wannier:1950},
\begin{equation}
S_{\text{W}}^{\circ}=\frac{2}{\pi}\intop_{0}^{\pi/3}\ln(2\cos\alpha)\,d\alpha\approx0.323\,066,\label{eq:S0:W}
\end{equation}
is exactly three times greater than that of the dual Calandre lattice
\[
\frac{S_{\text{W}}^{\circ}}{S^{\circ}}=3.
\]

\subsection{Calandre lattice particular cases}

To obtain and study numerous particular cases, in addition to duality relation two main methods can be used: zeroing one or more exchange interactions; striving exchange interactions to infinity. The last method had been proposed and developed in detail by Yamamoto \cite{Yamamoto:1951}. In the present paper, we will consider only a small number of special cases that already demonstrate the richness of the Calandre lattice.

\begin{figure}[tbh]
\centering \includegraphics{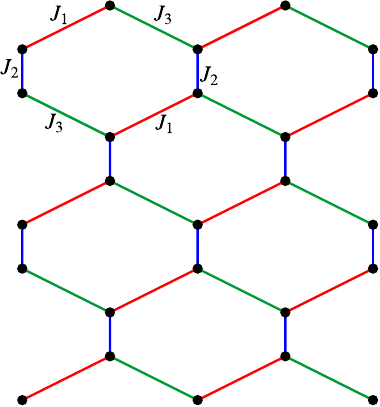}\qquad\includegraphics{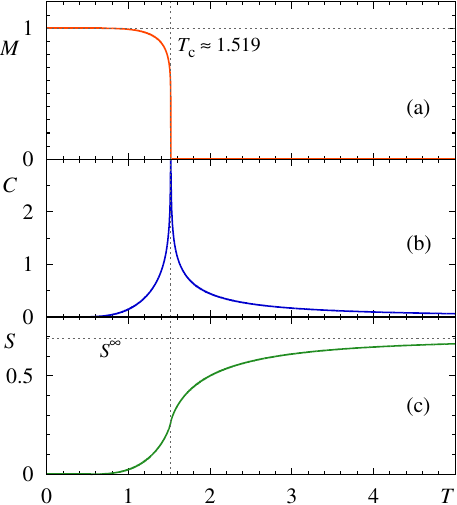}
\caption{The Calandre lattice with model parameters (\ref{eq:J:HEX}) (left panel). Temperature dependencies (right panel) of spontaneous magnetization $M(T)$ (a), specific heat $C(T)$ (b), and entropy $S(T)$ (c) for the case (\ref{eq:J:HEX}) for the Calandre lattice. The vertical dashed line indicates the position of the phase-transition temperature}
\label{fig:03}
\end{figure}

1) Let us consider the case where the fourth and fifth exchange interactions are equal to zero, while the first three interactions are ferromagnetic:
\begin{equation}
J_{1}=J_{2}=J_{3}=1,\quad J_{4}=J_{5}=0.\label{eq:J:HEX}
\end{equation}
As can be seen from Fig. \ref{fig:03}, this case corresponds to a hexagonal (honeycomb) lattice with a phase-transition temperature of 
\[
T_{\text{c}}=\frac{2}{\ln\left({\displaystyle 2+\sqrt{3}}\right)}\approx1.518\,651,
\]
and the phase-transition temperature of the lattice---the dual of the hexagonal lattice, i.e., the triangular lattice---
\[
T_{\text{c}}^{\text{dual}}=\frac{2}{\ln{\displaystyle \sqrt{3}}}\approx3.640\,957.
\]
The right panel of Fig. \ref{fig:03} shows the magnetic and thermodynamic functions for this lattice.

\begin{figure}[tbh]
\centering \includegraphics{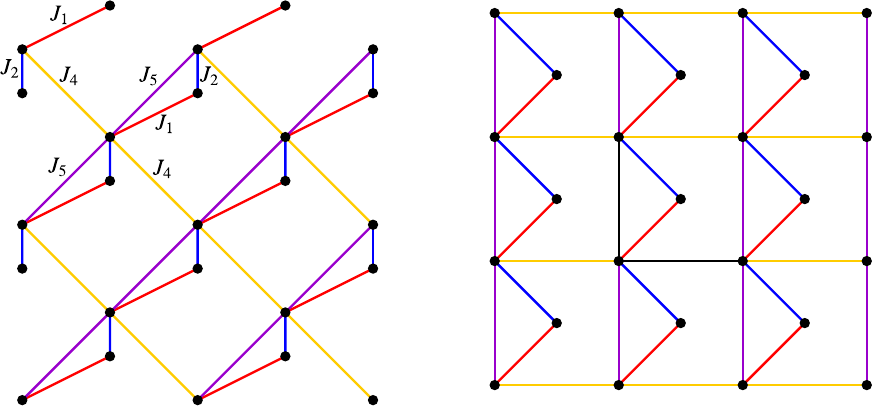}
\caption{The Calandre lattice with model parameters (\ref{eq:J:SDQ}) (left panel) and its topologically equivalent lattice (right panel)}
\label{fig:04}
\end{figure}

2) Let us consider the following case, where the first or third exchange interactions are equal to zero, while the remaining four interactions are ferromagnetic:
\begin{equation}
J_{1}=J_{2}=J_{4}=J_{5}=1,\quad J_{3}=0.\label{eq:J:SDQ}
\end{equation}
As can be seen from Fig. \ref{fig:04}, this case corresponds to a square lattice that has been decorated once in one of the directions. The right panel of Fig. \ref{fig:04} shows a lattice that is topologically equivalent to the one presented in the left panel.

\begin{figure}[tbh]
\centering \includegraphics{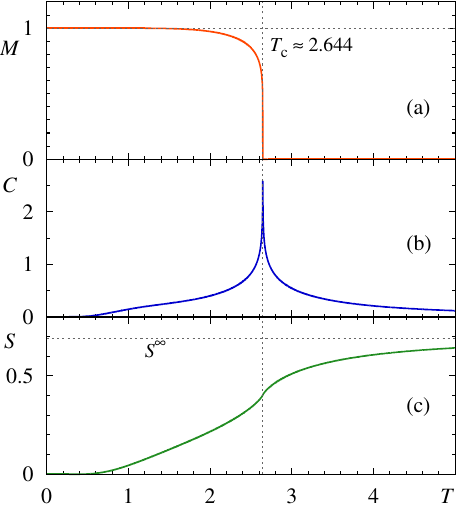}\qquad\includegraphics{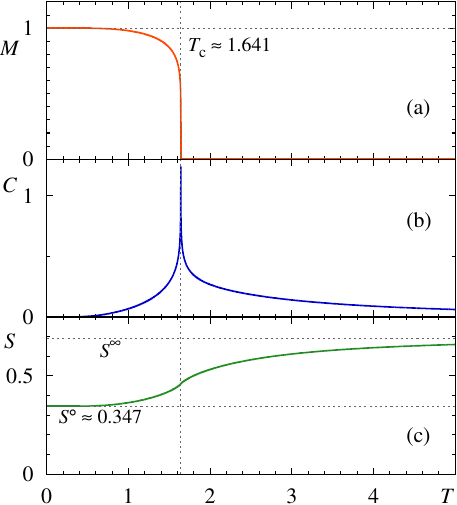}
\caption{Temperature dependencies of spontaneous magnetization $M(T)$ (a), specific heat $C(T)$ (b), and entropy $S(T)$ (c) for the case (\ref{eq:J:SDQ}) (left panel) and for the case (\ref{eq:J:SDQ:A}) (right panel) for the Calandre lattice. The vertical dashed line indicates the position of the phase-transition temperature, and on the right panel, the horizontal line on the figure (c), marked $S^{\circ}$, indicates the value of the residual entropy}
\label{fig:05}
\end{figure}

The phase-transition temperature in this case will be equal to 
\begin{equation}
T_{\text{c}}=\frac{2}{\ln\left({\displaystyle \frac{1}{3}+\frac{1}{3}\sqrt[3]{46+6\sqrt{57}}+\frac{1}{3}\sqrt[3]{46-6\sqrt{57}}}\right)}\approx2.644\,427.\label{eq:TC:2.644}
\end{equation}
The magnetic and thermodynamic functions of this lattice are presented in the left panel of Fig. \ref{fig:05}.

Note that if the ferromagnetic exchange interaction parameters in the set (\ref{eq:J:SDQ}) are replaced by antiferromagnetic ones
\begin{equation}
J_{1}=J_{2}=J_{4}=J_{5}=-1,\quad J_{3}=0,\label{eq:J:SDQ:A}
\end{equation}
a different phase-transition temperature, 
\begin{equation}
T_{\text{c}}=\frac{2}{\ln\left({\displaystyle 1+\frac{1}{3}\sqrt[3]{54+6\sqrt{33}}+\frac{1}{3}\sqrt[3]{54-6\sqrt{33}}}\right)}\approx1.641\,018,\label{eq:TC:1.641:}
\end{equation}
is obtained. Here it should be emphasized that we obtained the coincidence of the phase-transition points for two utterly dissimilar lattices and different exchange interaction signs.

In this process, a competition between exchange interactions arises within the system, leading to partial magnetic disorder and the emergence of magnetic frustration. It is precisely this magnetic frustration that gives rise to a residual entropy of 
\begin{equation}
S^{\circ}=\frac{\ln2}{2}\approx0.346\,574.\label{eq:S0}
\end{equation}
This behavior of the magnetic and thermodynamic functions is shown in the right panel of Fig. \ref{fig:05}, where it can be seen that the resulting magnetic frustrations ($S^{\circ}>0$) do not completely destroy the magnetic ordering in the spin system, and spontaneous magnetization persists at low temperatures.

Moreover, if one again makes use of the formulas derived in the paper \cite{Kassan-Ogly:2025}, which deals with a decorated square lattice, the result turns out to be the same. All this reveals a sudden connection between quite dissimilar lattices and with quite different exchange interactions.

\begin{figure}[tbh]
\centering \includegraphics{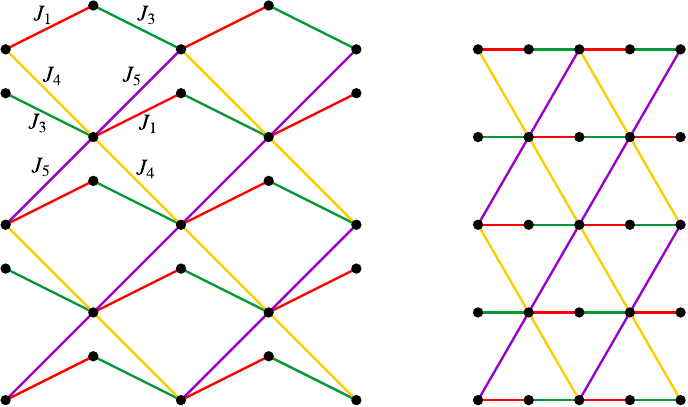}
\caption{The Calandre lattice with model parameters (\ref{eq:J:TRQ}) (left panel) and its topologically equivalent lattice (right panel)}
\label{fig:06}
\end{figure}

3) Consider the case when the second exchange interaction is zero, and the other four interactions are ferromagnetic:
\begin{equation}
J_{1}=J_{3}=J_{4}=J_{5}=1,\quad J_{2}=0.\label{eq:J:TRQ}
\end{equation}
As can be seen from Fig. \ref{fig:06}, for the given set of parameters, the lattice corresponds to a triangular lattice that is decorated once in one of the directions (see the right panel of the figure, which shows a topologically equivalent lattice).

\begin{figure}[tbh]
\centering \includegraphics{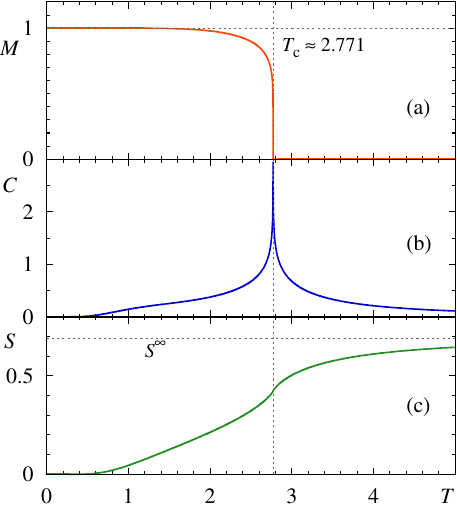}\qquad\includegraphics{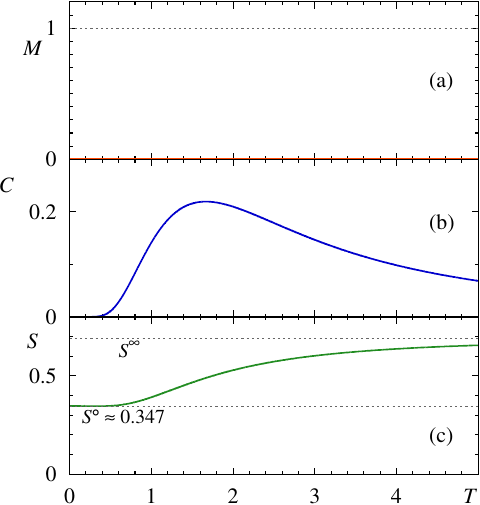}
\caption{Temperature dependencies of spontaneous magnetization $M(T)$ (a), specific heat $C(T)$ (b), and entropy $S(T)$ (c) for the case (\ref{eq:J:TRQ}) (left panel) and for the case (\ref{eq:J:TRQ:A}) (right panel) for the Calandre lattice. The vertical dashed line in the left panel indicates the position of the phase-transition temperature, and on the right panel, the horizontal line on the figure (c), marked $S^{\circ}$, indicates the value of the residual entropy}
\label{fig:07}
\end{figure}

In this case, the phase-transition temperature is 
\begin{equation}
T_{\text{c}}=\frac{2}{\ln\left({\displaystyle \sqrt{2+\sqrt{5}}}\right)}\approx2.770\,783,\label{eq:TC:2.771}
\end{equation}
as can be seen in the left panel of Fig. \ref{fig:07}, which displays the magnetic and thermodynamic functions for this lattice.

In addition, if we now use the formulas obtained in the article \cite{Kassan-Ogly:2023}, devoted to a decorated triangular lattice, we obtain the exact same expression for the phase-transition point.

It is also important to note that if any one or three of the ferromagnetic exchange interaction parameters in the set (\ref{eq:J:TRQ}) are replaced by antiferromagnetic ones, for example, by setting 
\begin{equation}
J_{1}=-1,\quad J_{3}=J_{4}=J_{5}=1,\quad J_{2}=0,\label{eq:J:TRQ:A}
\end{equation}
the competition between exchange interactions will inevitably arise within the spin system. This competition gives rise to a specific type of magnetic frustration, which suppresses magnetic ordering and results in the absence of spontaneous magnetization. In this case, the residual entropy of the system is given by (\ref{eq:S0}). This behavior of the magnetic and thermodynamic functions is clearly illustrated in the right panel of Fig. \ref{fig:07}.

4) Now let us turn to the joint consideration of the case that encompasses four distinct lattices and therefore merits separate examination.

\begin{figure}[tbh]
\centering \includegraphics{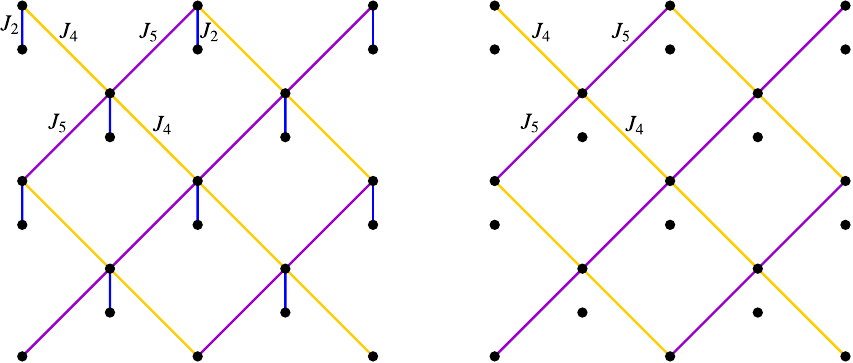}
\caption{The Calandre lattice with model parameters (\ref{eq:J:DIMER}) (left panel) and with model parameters (\ref{eq:J:SQF}) (right panel)}
\label{fig:08}
\end{figure}

4.1) The first variant is characterized by the following set of model parameters for
\begin{equation}
J_{2}=J_{4}=J_{5}=1,\quad J_{1}=J_{3}=0,\label{eq:J:DIMER}
\end{equation}
The lattice with these parameters is shown in Fig. \ref{fig:08} (left panel). In this case, it can be seen that a dimer is located at one of the sites of the square lattice (marked in blue in the figure).

4.2) The second variant is defined by the parameter set 
\begin{equation}
J_{4}=J_{5}=1,\quad J_{1}=J_{2}=J_{3}=0.\label{eq:J:SQF}
\end{equation}
In this case, the square lattice contains, within each square, a site with a “free” spin---that is, a spin that is not coupled via exchange interaction to any spin at neighboring sites. This case is shown in the right panel of Fig. \ref{fig:08}.

4.3) The third variant has the following set of parameters: 
\begin{equation}
J_{1}=J_{3}=1,\quad J_{2}=\infty,\quad J_{4}=J_{5}=0.\label{eq:J:SQ0}
\end{equation}
Here, we have a simple square lattice.

\begin{figure}[tbh]
\centering \includegraphics{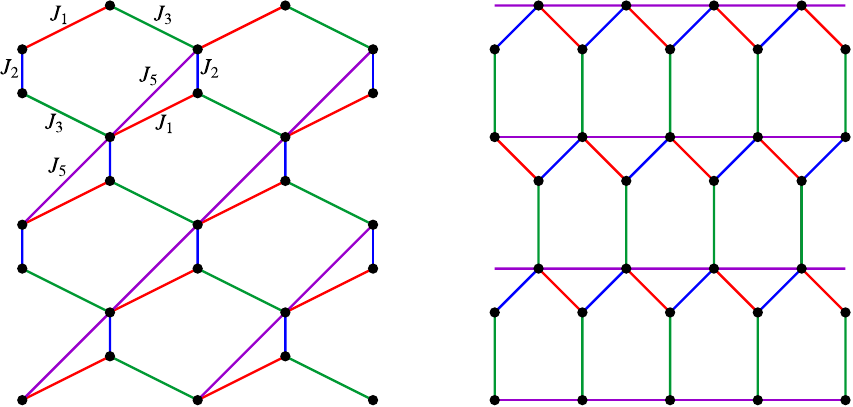}
\caption{The Calandre lattice with model parameters (\ref{eq:J:SDUAL}) (left panel) and its topologically equivalent lattice (right panel)}
\label{fig:09}
\end{figure}

4.4) The fourth variant is defined by the parameter set 
\begin{equation}
J_{1}=J_{2}=J_{3}=J_{5}=1,\quad J_{4}=0.\label{eq:J:SDUAL}
\end{equation}
In this case, we obtain the lattice shown in Fig. \ref{fig:09}. This is the self-dual 3-5 lattice from \cite{Chao:1990}, which is depicted in the right panel of Fig. \ref{fig:09}.

Thus, four different types of lattices are presented above, each originating from a single source --- the Calandre lattice. Despite their outward dissimilarity, these distinct lattices share one common parameter: a phase-transition temperature of 
\[
T_{\text{c}}=\frac{2}{\ln\left(1+\sqrt{2}\right)}\approx2.269\,185.
\]
This is the famous Kramers--Wannier phase-transition point for the square lattice, which had been obtained by the duality relatioin.

\begin{figure}[tbh]
\centering \includegraphics{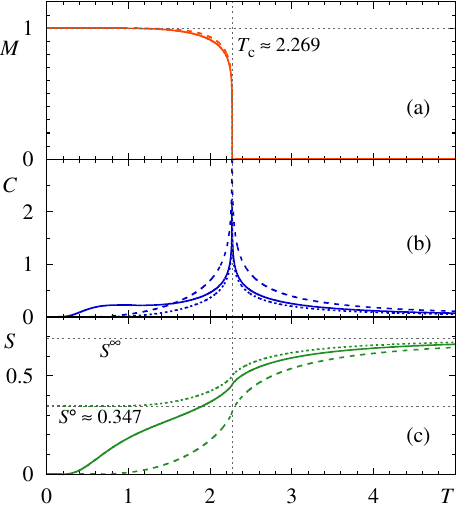}\qquad\includegraphics{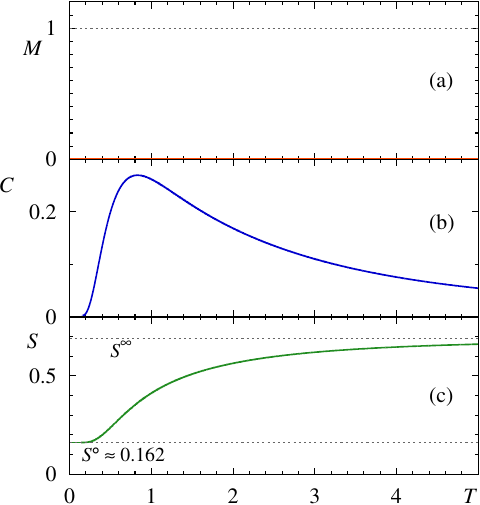}
\caption{Temperature dependencies of spontaneous magnetization $M(T)$ (a), specific heat $C(T)$ (b), and entropy $S(T)$ (c). In the left panel, the solid line corresponds to the model parameters (\ref{eq:J:DIMER}), the dotted line to (\ref{eq:J:SQF}), and the dashed line to (\ref{eq:J:SDUAL}). The right panel shows the temperature dependencies of the functions for the model parameters (\ref{eq:J:SDUAL:A}). The vertical dashed line in the left panel indicates the position of the phase-transition temperature, and the horizontal line on the figure (c), marked $S^{\circ}$, indicates the value of the residual entropy}
\label{fig:10}
\end{figure}

The magnetic and thermodynamic functions for these lattices are shown in the left panel of Fig. \ref{fig:10}.

It is important to note that the lattice corresponding to the set of parameters (\ref{eq:J:SQF})---shown in the right panel of Fig. \ref{fig:08}---contains “free” spins. These free spins are precisely what determine the degeneracy of the ground state, which, in turn, gives rise to a residual entropy, the value of which is given by the expression (\ref{eq:S0}). The corresponding value of the residual entropy, $S^{\circ}$, is indicated by a dashed line in Fig. \ref{fig:10}(c).

Finally, it should be noted that for the lattice shown in Fig. \ref{fig:09}, replacing the ferromagnetic exchange interaction parameters in set (\ref{eq:J:SDUAL}) with antiferromagnetic ones 
\begin{equation}
J_{1}=J_{2}=J_{3}=J_{5}=-1,\quad J_{4}=0.\label{eq:J:SDUAL:A}
\end{equation}
inevitably gives rise to competing exchange interactions within the spin system (a situation analogous to that shown in the right panel of Fig. \ref{fig:07}). This competition leads to the emergence of magnetic frustrations that suppress magnetic ordering. Consequently, a residual entropy is formed, and there is no spontaneous magnetization. In this case, the residual entropy of the system is 
\begin{equation}
S_{\text{W}}^{\circ}=\frac{1}{\pi}\intop_{0}^{\pi/3}\ln(2\cos\alpha)\,d\alpha\approx0.161\,533,\label{eq:S0:0.162}
\end{equation}
which is half the value of the Wannier entropy (\ref{eq:S0:W}). This behavior of the magnetic and thermodynamic functions is illustrated in the right panel of Fig. \ref{fig:10}.

\subsection{Magnetic reentrancy on the Calandre lattice}

In conclusion, let us consider the last, but perhaps very important case, which demonstrates the fundamental possibility of the occurrence of multiple phase transitions in the Calandre lattice. To be sure, multiple phase transitions (in other words, magnetic reentrancy) have indeed been discovered in planar lattices \cite{Zarubin:2026}; however, in all known cases we are talking about either decorated lattices or lattices in which interactions not only between nearest neighbors are taken into account. It is this circumstance that gives the example under consideration special significance: it shows that for the emergence of magnetic reentrancy it is not at all necessary to resort to such complications of the structure or long-range action.

\begin{figure}[tbh]
\centering \includegraphics{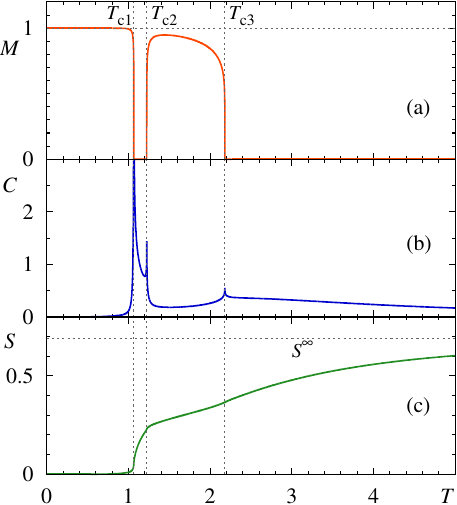}
\caption{Temperature dependencies of spontaneous magnetization $M(T)$ (a), specific heat $C(T)$ (b), and entropy $S(T)$ (c) for the parameters (\ref{eq:J:3TC}). The vertical dashed lines indicate the positions of the phase-transition temperatures $T_{\text{c}1}\approx1.065$, $T_{\text{c}2}\approx1.225$, and $T_{\text{c}3}\approx2.179$}
\label{fig:11}
\end{figure}

Let us focus on one of many possible scenarios: the model with a parameter set of 
\begin{equation}
J_{1}=J_{2}=J_{5}=1,\quad J_{3}=-1.1,\quad J_{4}=2.2\label{eq:J:3TC}
\end{equation}
for the Calandre lattice, shown in Fig. \ref{fig:01}. As illustrated in Fig. \ref{fig:11}, this system undergoes three magnetic phase transitions, which occur at temperatures $T_{\text{c}1}\approx1.065$, $T_{\text{c}2}\approx1.225$, and $T_{\text{c}3}\approx2.179$. These transitions are clearly manifested in both the temperature dependence of the spontaneous magnetization (Fig. \ref{fig:11}(a)) and the temperature dependence of the specific heat (Fig. \ref{fig:11}(b)); each curve exhibits characteristic features corresponding to the aforementioned temperatures. It is significant that the system exhibits neither magnetic degeneracy nor magnetic frustration, a fact confirmed by the entropy vanishing at zero temperature (Fig. \ref{fig:11}(c)).

\section{Conclusions}

In the present work, an exact solution for the Helmholtz free energy in the Ising model on the planar Calandre lattice has been obtained in a canonical form through the principal eigenvalue of the Kramers--Wannier transfer-matrix. Furthermore, an expression for spontaneous magnetization has been derived, as well as a general formula connecting the phase-transition points in any pair of the original and dual lattices.

Using these formulas, the thermodynamic, magnetic, and frustration properties of both the model itself and its various special cases have been investigated.

For all phase-transition points, as well as for residual entropy (in the presence of magnetic frustrations in the system), exact analytical expressions have been derived.

It has been established that the Calandre lattice possesses some unique properties inherent only to it and not observed in any other lattices. In particular, it turned out that the ratio of Curie points in the Calandre lattice and its dual is exactly equal to two, and the residual entropy in the dual Calandre lattice is exactly three times less than that in the triangular lattice \cite{Wannier:1950}. Moreover, a paradoxical fact has been revealed, namely, that the Calandre lattice, despite being \textit{undecorated}, exhibits connections with known \textit{decorated} lattices in special cases.

It should be noted that throughout the article, significant restrictions have been imposed on the exchange integrals, namely, in absolute value, they are always equal to either one or zero, as is customary in such calculations within statistical physics.

And only one example is provided where these restrictions are lifted, once again revealing the uniqueness of the Calandre lattice (Fig. \ref{fig:11}). The figure clearly shows the phenomenon of multiple phase transitions and the reentrant behavior, which are typically characteristic of decorated lattices \cite{Miyazima:1968}.

Despite the multitude of results obtained, we believe that the Calandre lattice, both when the restrictions on exchange integrals are lifted and when direct consideration of decorations is taken into account, holds considerable potential for revealing new phenomena and effects, deserving close attention and requiring further, deeper investigations. 


\section*{Acknowledgments}

The work was carried out within the framework of the state assignment of the Ministry of Science and Higher Education of the Russian Federation for the IMP UB RAS.

\bibliographystyle{apsrev4-2}
\bibliography{calandre}

\end{document}